\documentclass[conference]{IEEEtran}
\input{psfig.sty}
\usepackage{subfigure}
\usepackage{multirow}
\usepackage{psfig}
\usepackage{graphicx}
\usepackage{amsmath}
\usepackage{color}
\usepackage{epsfig}
\usepackage{amsfonts}
\usepackage{amssymb}
\usepackage{amsthm}
\usepackage{algorithm2e}
\usepackage{algorithmic}
\usepackage[usenames,dvipsnames]{pstricks}

\newcommand{\argmax}{\mathop{\text{argmax}}}
\newcommand{\argmin}{\mathop{\text{argmin}}}

\newcommand{\vh}{{\bf h}}
\newcommand{\vw}{{\bf w}}
\newcommand{\vz}{{\bf z}}
\newcommand{\vx}{{\bf x}}
\newcommand{\vy}{{\bf y}}
\newcommand{\vs}{{\bf s}}
\newcommand{\vn}{{\bf n}}

\newcommand{\mh}{{\bf H}}

\newcommand{\sm}{{\mathbb S}_{n_t,{\mathbb A}}}
\newcommand{\sa}{{\mathbb A}}
\newcommand{\E}{{\mathbb E}}

\newcommand{\Define}{\triangleq}

\begin{document}
\twocolumn

\title{{\huge 
Multiuser SM-MIMO versus Massive MIMO: Uplink Performance Comparison
}}

\author{
P. Raviteja, T. Lakshmi Narasimhan and A. Chockalingam \\
Department of ECE, Indian Institute of Science, Bangalore}
\IEEEaftertitletext{\vspace{-0.6\baselineskip}}
\maketitle

\begin{abstract}
In this paper, we propose algorithms for signal detection in large-scale 
multiuser {\em spatial modulation multiple-input multiple-output (SM-MIMO)} 
systems. In large-scale SM-MIMO, each user is equipped with multiple transmit 
antennas (e.g., 2 or 4 antennas) but only one transmit RF chain, and the base 
station (BS) is equipped with tens to hundreds of (e.g., 128) receive antennas. 
In SM-MIMO, in a given channel use, each user activates any one of its 
multiple transmit antennas and the index of the activated antenna conveys 
information bits in addition to the information bits conveyed through 
conventional modulation symbols (e.g., QAM). We propose two different 
algorithms for detection of large-scale SM-MIMO signals at the BS; one is 
based on {\em message passing} and the other is based on {\em local search}.
The proposed algorithms are shown to achieve very good performance and scale
well. Also, for the same spectral efficiency, multiuser SM-MIMO 
outperforms conventional multiuser MIMO (recently being referred to as 
massive MIMO) by several dBs; for e.g., with 16 users, 128 antennas at the 
BS and 4 bpcu per user, SM-MIMO with 4 transmit antennas per user and 4-QAM 
outperforms massive MIMO with 1 transmit antenna per user and 16-QAM by 
about 4 to 5 dB at $10^{-3}$ uncoded BER. The SNR advantage of SM-MIMO
over massive MIMO can be attributed to the following reasons: $(i)$ 
because of the spatial index bits, SM-MIMO can use a lower-order QAM
alphabet compared to that in massive MIMO to achieve the same spectral
efficiency, and $(ii)$ for the same spectral efficiency and QAM size,
massive MIMO will need more spatial streams per user which leads to 
increased spatial interference.
\end{abstract}
\vspace{-0.5mm}

\begin{figure*}
\subfigure[SM-MIMO system.]{
\includegraphics[width=3.15in,height=3.00in]{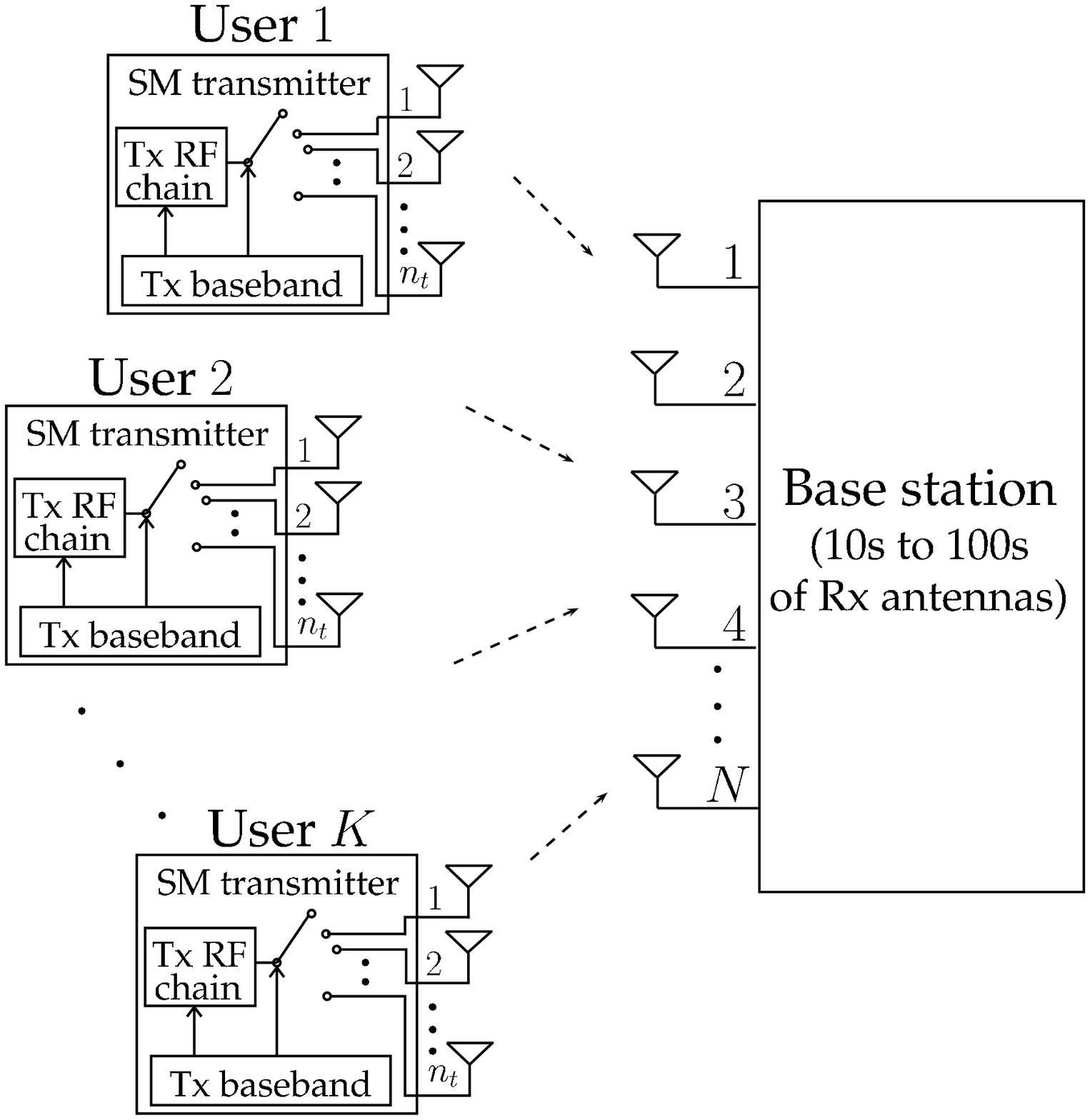}
\label{sys1}
}
\hspace{6mm}
\subfigure[Massive MIMO system.]{
\includegraphics[width=3.15in,height=3.00in]{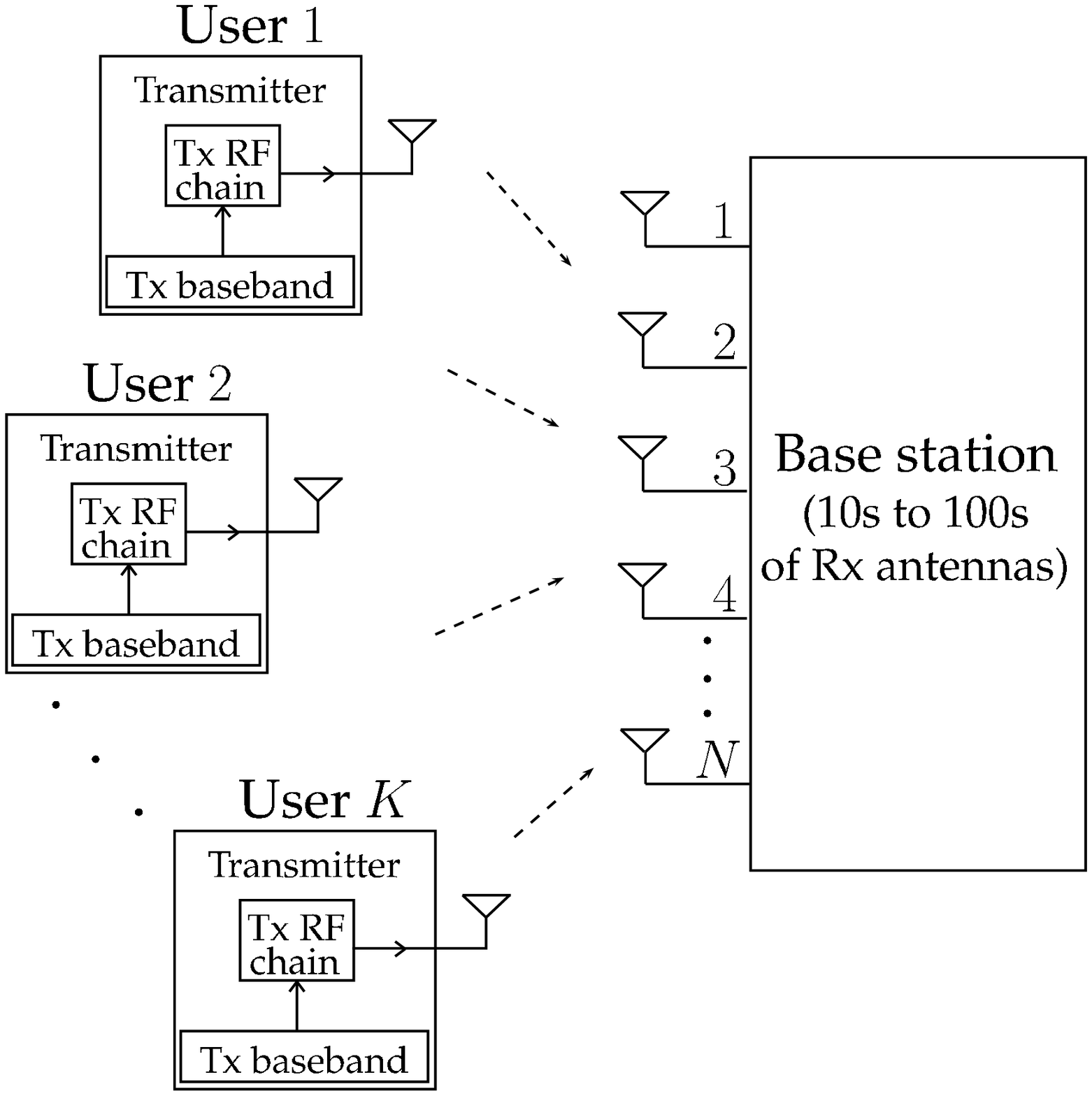}
\label{sys2}
}
\caption{Large-scale multiuser SM-MIMO and massive MIMO system architectures.}
\vspace{-4mm}
\end{figure*}

\section{Introduction}
\label{sec1}
Large-scale MIMO systems with tens to hundreds of antennas are getting
increased research attention \cite{lasx}-\cite{mmse1}. 
The following two characteristics are typical in conventional MIMO systems: 
$(i)$ there will be one transmit RF chain for each transmit antenna (i.e., 
on the modulation symbols (e.g., QAM). Spatial modulation MIMO (SM-MIMO) 
systems \cite{SM_commag} differ from conventional MIMO systems in the 
following two aspects: $(i)$ in SM-MIMO there will be multiple transmit 
antennas but only one transmit RF chain, and $(ii)$ the index of the active 
transmit antenna will also convey information bits in addition to information 
bits conveyed through modulation symbols like QAM. The advantages of 
SM-MIMO include reduced RF hardware complexity, size, and cost. 

Conventional multiuser MIMO systems with a large number (tens to hundreds)
of antennas at the base station (BS) are  referred to as `massive MIMO' 
systems in the recent literature \cite{mmse1}. The users in a massive MIMO 
system can have one or more transmit antennas with equal number of transmit 
RF chains. In large-scale multiuser SM-MIMO systems also, the number of BS
antennas will be large. The users in SM-MIMO will have multiple transmit 
antennas but only on RF chain. Figures \ref{sys1} and \ref{sys2} illustrate 
the large-scale multiuser SM-MIMO system (with $K$ users, $N$ BS antennas,
$n_t$ transmit antennas per user, and $n_{rf}=1$ transmit RF chain per user) 
and massive MIMO system (with $K$ users, $N$ BS antennas, $n_t=1$ transmit 
antenna per user, and $n_{rf}=1$ transmit RF chains per user), respectively.

Several works have focused on single user point-to-point SM-MIMO systems 
(\cite{lajos} and the references therein). Some works on multiuser SM-MIMO 
have also been reported \cite{mu1}-\cite{mu4}. An interesting result reported 
in \cite{mu1} is that multiuser SM-MIMO outperforms conventional multiuser MIMO 
by several dBs for the same spectral efficiency. This work is limited to 3 users 
(with 4 antennas each) and 4 antennas at BS receiver. Also, only maximum 
likelihood (ML) detection is considered. This superiority of SM-MIMO over 
conventional MIMO attracts further investigations on multiuser SM-MIMO. 
In particular, investigations in the following two directions are of interest:
$(i)$ large-scale SM-MIMO (with large number of users and BS antennas), and
$(ii)$ detection algorithms that can scale and perform well in such large-scale
SM-MIMO systems. In this paper, we make contributions in these two directions.

We investigate multiuser SM-MIMO with similar number of users and BS antennas
envisaged in massive MIMO, e.g., tens of users and hundreds of BS antennas.
Our contributions can be summarized as follows.
\begin{itemize}
\vspace{-1mm}
\item Proposal of two different algorithms for detection of large-scale SM-MIMO 
      signals at the BS. One algorithm is based on {\em message passing} 
      referred to as MPD-SM (message passing detection for spatial modulation) 
      algorithm, and the other is based on {\em local search} referred to as
      LSD-SM (local search detection for spatial modulation) algorithm.
      Simulation results show that these proposed algorithms achieve very good 
      performance and scale well. 
\item Uplink performance comparison between SM-MIMO and massive MIMO for the 
      same spectral efficiency. Simulation results show that SM-MIMO outperforms 
      massive MIMO by several dBs; e.g., SM-MIMO has a 4 to 5 dB SNR advantage 
      over massive MIMO at $10^{-3}$ BER for 16 users, 128 BS antennas, 
      and 4 bpcu per user. 

      The SNR advantage of SM-MIMO over massive MIMO is 
      attributed to the following reasons: $(i)$ because of the spatial index 
      bits, SM-MIMO can use a lower-order QAM alphabet compared to that in 
      massive MIMO to achieve the same spectral efficiency, and $(ii)$ for the 
      same spectral efficiency and QAM size, massive MIMO will need more spatial 
      streams per user which leads to increased spatial interference.
\end{itemize}

The rest of the paper is organized as follows. The system model for
multiuser SM-MIMO is presented in Section \ref{sec2}. The proposed
MPD-SM algorithm for detection of SM-MIMO signals and its performance
are presented in Section \ref{sec3}. In Section \ref{sec4}, the proposed 
LSD-SM algorithm and its performance are presented. Performance comparison
between SM-MIMO and massive MIMO is presented in Sections \ref{sec3}
and \ref{sec4}. Conclusions are presented in Section \ref{sec5}.

\section{Multiuser SM-MIMO system model}
\label{sec2}
Consider a multiuser system with $K$ uplink users communicating with 
a BS having $N$ receive antennas, where $N$ is in the order of tens 
to hundreds. The ratio $\alpha=K/N$ is the system loading factor. 
Each user employs spatial modulation (SM) for transmission, where 
each user has $n_t$ transmit antennas but only one transmit RF chain 
(see Fig. \ref{sys1}). In a given channel use, each user selects any one 
of its $n_t$ transmit antennas, and transmits a symbol from a modulation 
alphabet ${\mathbb A}$ on the selected antenna. The number of bits 
conveyed per channel use per user through the modulation symbols is 
$\lfloor \log_2|{\mathbb A}| \rfloor$. In addition, 
$\lfloor \log_2n_t \rfloor$ bits per channel use (bpcu) per user is 
conveyed through the index of the chosen transmit antenna. Therefore, 
the overall system throughput is 
$K(\lfloor \log_2|{\mathbb A}|\rfloor + \lfloor \log_2 n_t \rfloor)$ bpcu. 
For e.g., in a system with $K=3$, $n_t=4$, 4-QAM, the system 
throughput is 12 bpcu. 

The SM signal set ${\mathbb S}_{n_t,{\mathbb A}}$ for each user is given by 
\begin{eqnarray}
{\mathbb S}_{n_t,{\mathbb A}} = 
\big \{ {\bf s}_{j,l}:j=1,\cdots,n_t, \ \ l=1,\cdots,|{\mathbb A}| \big \}, \nonumber \\ 
\mbox{s.t.} \ \ {\bf s}_{j,l} = 
[0,\cdots,0,\hspace{-4mm}\underbrace{s_{l}}_{{\scriptsize{\mbox{$j$th coordinate}}}}\hspace{-3.5mm},0,\cdots,0]^T, \ \ s_l \in \mathbb{A}.  
\end{eqnarray}
For e.g., for $n_t=2$ and 4-QAM, ${\mathbb S}_{n_t,{\mathbb A}}$ is given by  
{\small
\begin{eqnarray}
\hspace{-4mm}
{\mathbb S}_{2,\mbox{{\tiny 4-QAM}}}  
\hspace{-3mm}&=&\hspace{-3mm}\Bigg\{ 
\begin{bmatrix} +1+j \\ 0 \end{bmatrix}, 
\begin{bmatrix} +1-j \\ 0 \end{bmatrix}, 
\begin{bmatrix} -1+j \\ 0 \end{bmatrix}, 
\begin{bmatrix} -1-j \\ 0 \end{bmatrix}, \nonumber \\ 
& & 
\begin{bmatrix} 0 \\ +1+j \end{bmatrix}, 
\begin{bmatrix} 0 \\ +1-j \end{bmatrix}, 
\begin{bmatrix} 0 \\ -1+j \end{bmatrix}, 
\begin{bmatrix} 0 \\ -1-j \end{bmatrix} 
\Bigg \}. 
\end{eqnarray}
}

\vspace{-3mm}
Let $\vx_k \in {\mathbb S}_{n_t,{\mathbb A}}$ denote the transmit vector from 
user $k$. Let
$\vx \Define [\vx_1^T \ \ \vx_2^T\,\cdots\,\vx_k^T\,\cdots\,\vx_K^T]^T$ 
denote the vector comprising of transmit vectors from all the users. 
Note that $\vx \in \sm^K$.

Let $\mh \in \mathbb{C}^{N\times Kn_t}$ denote the channel gain matrix,
where $H_{i,(k-1)n_t+j}$ denotes the complex channel gain from the $j$th 
transmit antenna of the $k$th user to the $i$th BS receive antenna. The 
channel gains are assumed to be independent Gaussian with zero mean and 
variance $\sigma_k^2$, such that $\sum_k \sigma_k^2=K$. The $\sigma_k^2$ 
models the imbalance in the received power from user $k$ due to path loss 
etc., and $\sigma_k^2=1$ corresponds to the case of perfect power control. 
Assuming perfect synchronization, the received signal at the $i$th BS
antenna is given by
\begin{equation}
y_i = \sum_{k=1}^{K} x_{l_k}H_{i,(k-1)n_t+j_k} + n_i,
\end{equation}
where $x_{l_k}$ is the $l_k$th symbol in ${\mathbb A}$, transmitted by the 
$j_k$th antenna of the $k$th user, and $n_i$ is the noise modeled as a complex 
Gaussian random variable with zero mean and variance $\sigma^2$. The received 
signal at the BS antennas can be written in vector form as
\begin{eqnarray}
\vy& = & \mh\vx+\vn,
\label{sysmodel}
\end{eqnarray}
where $\vy = [y_1, \ y_2, \cdots, \ y_N]^T$ and 
$\vn = [n_1, \ n_2, \cdots, \ n_N]^T$. 

For this system model, the maximum-likelihood (ML) detection rule is 
given by
\begin{equation} 
\label{ml} 
\hat{\vx}=\argmin_{\vx\in \sm^{K}} \ \|\vy-\mh\vx\|^2,  
\end{equation}
where $\|\vy-\mh\vx\|^2$ is the ML cost.
The maximum a posteriori  probability (MAP) decision rule, is given by
\begin{eqnarray}
\label{map}
\hat{\vx} = \argmax_{\vx\in \sm^{K}} \ \Pr(\vx\mid\vy,\mh). 
\end{eqnarray}
Since $|\sm^{K}|=(|\sa|n_t)^K$,
the exact computation of (\ref{ml}) and (\ref{map}) requires exponential
complexity in $K$. We propose two low complexity detection algorithms for
multiuser SM-MIMO; one based on message passing (Sec. \ref{sec3}) which 
gives an approximate solution to (\ref{map}), and another based on local search 
(Sec. \ref{sec4}) which gives an approximate solution to (\ref{ml}). 

Note that in conventional multiuser MIMO, the vector $\vx$ in (\ref{sysmodel}) 
is $\vx \in {\mathbb B}^K$ where ${\mathbb B}$ is the modulation alphabet, and
$\mh \in {\mathbb C}^{N\times K}$. The condition for SM-MIMO and conventional 
MIMO to have the same system throughput is $|{\mathbb B}| = |\sa|n_t$. 

\section{Message Passing Detection for SM-MIMO}
\label{sec3}
\begin{figure}
\centering 
\subfigure[Factor graph]{
\includegraphics[width=2.75in,height=1.25in]{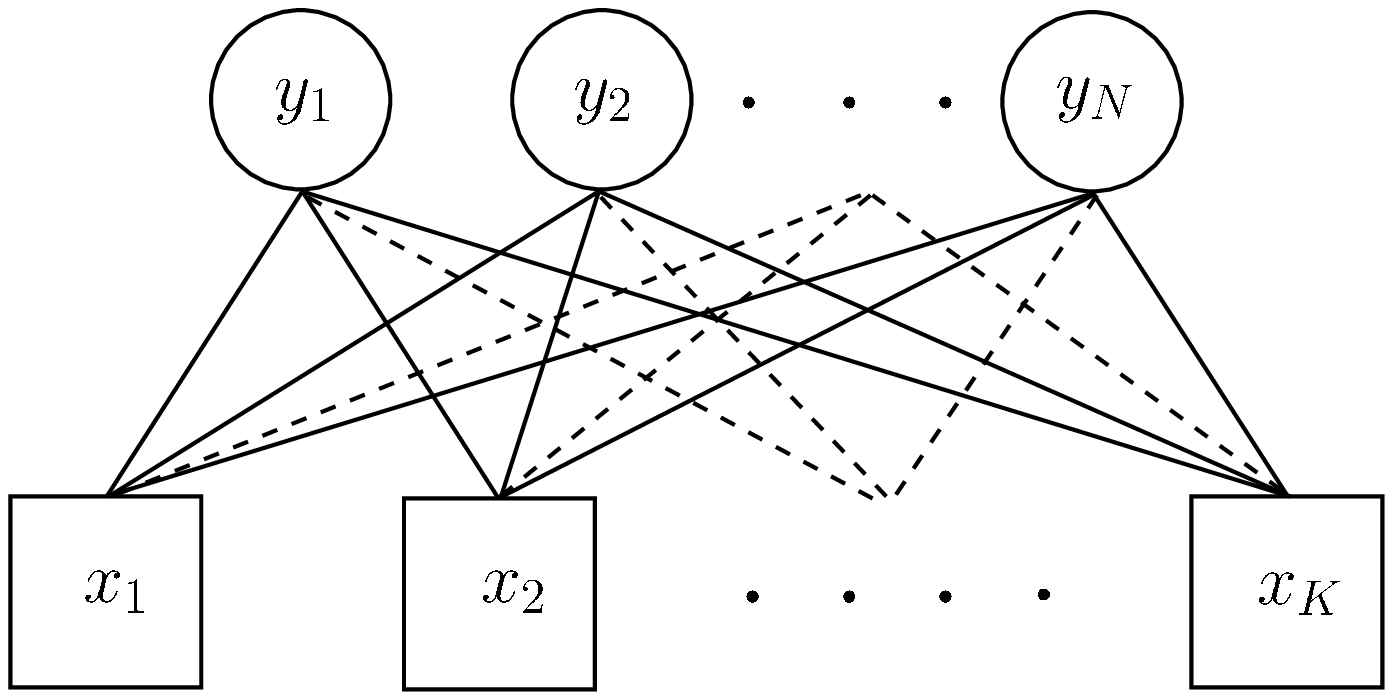}
\label{graph}
}
\subfigure[Observation node messages]{
\includegraphics[width=1.3in,height=1in]{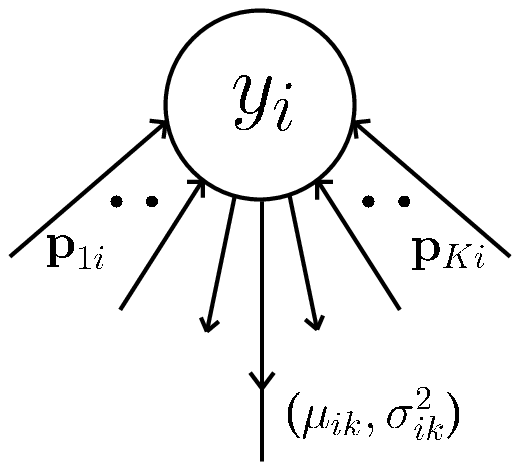}
\label{obs}
}
\subfigure[Variable node messages]{
\hspace{4mm}
\includegraphics[width=1.5in,height=0.9in]{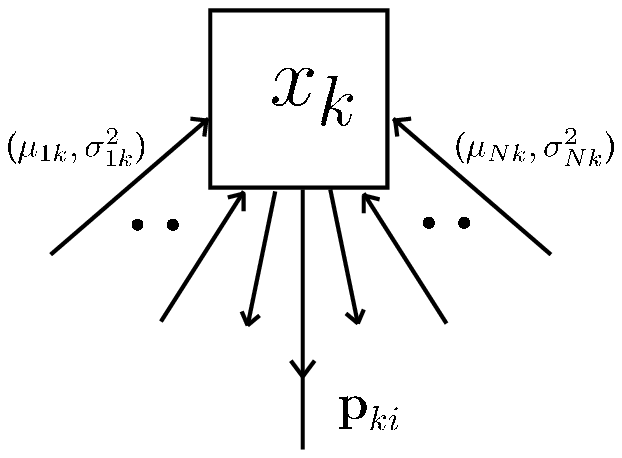}
\label{var}
}
\caption{The factor graph and messages passed in MPD-SM algorithm.}
\vspace{-2mm}
\end{figure}

In this section, we propose a message passing based algorithm for detection 
in SM-MIMO systems. We refer to the proposed algorithm as the MPD-SM
(message passing detection for spatial modulation) algorithm.
We model the system as a fully connected factor graph with $K$ variable 
(or factor) nodes corresponding to $\vx_k$'s and $N$ observation nodes 
corresponding to $y_i$'s, as shown in Fig. \ref{graph}.

{\em Messages:}
We derive the messages passed in the factor graph as follows.
Equation (\ref{sysmodel}) can be written as
\begin{equation}
y_i= {\bf h}_{i,[k]}\vx_k+
\underbrace{\sum_{j=1, j\neq k}^K{\bf h}_{i,[j]}\vx_j + n_i}_{\Define \ g_{ik}},
\label{usreqn}
\end{equation}
where ${\bf h}_{i,[j]}$ is a row vector of length $n_t$, given by 
$[H_{i,(j-1)n_t+1} \quad H_{i,(j-1)n_t+2} \, \cdots \, H_{i,jn_t}]$, 
and $\vx_j \in \sm$.

We approximate the term $g_{ik}$ to have a Gaussian distribution\footnote{This 
Gaussian approximation will be accurate for large $K$; e.g., in systems with 
tens of users.} with mean $\mu_{ik}$ and variance $\sigma_{ik}^2$ as follows. 

\vspace{-4mm}
{\small
\begin{eqnarray}
\mu_{ik}&\hspace{-2.5mm} = & \hspace{-2.5mm} \E\bigg[\sum_{j=1, j\neq k}^K \hspace{-2mm} {\bf h}_{i,[j]}\vx_j + n_i\bigg] 
        =\sum_{j=1, j\neq k}^K \ \sum_{\vs\in\sm} \hspace{-2mm} p_{ji}(\vs){\bf h}_{i,[j]}\vs
            \nonumber\\
&=&\sum_{j=1, j\neq k}^K \ \sum_{\vs\in\sm}p_{ji}(\vs) s_{l_s} H_{i,(j-1)n_t+l_s},
\label{mu}
\end{eqnarray} 
}

\vspace{-3mm}
where $s_{l_s}$ is the only non-zero entry in $\vs$ and $l_s$ is its index,
and $p_{ki}(\vs)$ is the message from $k$th variable node to the $i$th 
observation node. The variance is given by

\vspace{-4mm}
{\small
\begin{eqnarray}
\sigma_{ik}^2&=&\text{Var}\bigg(\sum_{j=1, j\neq k}^K{\bf h}_{i,[j]}\vx_j+n_i\bigg) 
			\nonumber\\
&=&\sum_{j=1, j\neq k}^K \ \sum_{\vs\in\sm}
p_{ji}(\vs){\bf h}_{i,[j]} \vs \vs^H {\bf h}_{i,[j]}^H \nonumber\\
&& -\Big|\sum_{\vs\in\sm}p_{ji}(\vs) {\bf h}_{i,[j]}\vs\Big|^2 +\sigma^2
		\nonumber\\
&=&\sum_{j=1, j\neq k}^K \ \sum_{\vs\in\sm}
p_{ji}(\vs)\big|s_{l_s}H_{i,(j-1)n_t+l_s}\big|^2\nonumber\\
&&\quad -\Big|\sum_{\vs\in\sm}p_{ji}(\vs)s_{l_s}H_{i,(j-1)n_t+l_s}\Big|^2 +\sigma^2.
\label{sigma}
\end{eqnarray} 
}

\vspace{-2mm}
The message $p_{ki}(\vs)$ is given by
{\small
\begin{equation}
p_{ki}(\vs)\propto\prod_{m=1, m\neq i}^N\exp\Big(-
	\frac{\big|y_m-\mu_{mk}-{\bf h}_{m,[k]}\vs\big|^2}{2\sigma^2_{mk}}\Big).
\label{pki}
\end{equation}
}

{\em Message passing:}
The message passing is done as follows.

{\bfseries {\em Step 1}}: Initialize $p_{ki}(\vs)$ to $1/|\sm|$ for all $i$, $k$ and $\vs$.\\
{\bfseries {\em Step 2}}: Compute $\mu_{ik}$ and $\sigma^2_{ik}$ from (\ref{mu}) and
	      (\ref{sigma}), respectively.\\
{\bfseries {\em Step 3}}: Compute $p_{ki}$ from (\ref{pki}). To improve the 
               convergence 
	      rate, damping \cite{damp} of the messages in (\ref{pki}) is done 
	      with a damping factor $\delta\in (0,1]$.\\
Repeat Steps 2 and 3 for a certain number of iterations. 
Figures \ref{obs} and \ref{var} illustrate the exchange of messages between
observation and variable nodes, where the vector message
${\bf p}_{ki}=[p_{ki}(\vs_1), p_{ki}(\vs_2),\cdots,p_{ki}(\vs_{|\sm|})]$. 
The final symbol probabilities at the end 
are given by
\begin{equation}
p_{k}(\vs)\propto\prod_{m=1}^N\exp\Big(-
\frac{\big|y_m-\mu_{mk}-{\bf h}_{m,[k]}\vs\big|^2}{2\sigma^2_{mk}}\Big).
\label{probs}
\end{equation}
The detected vector of the $k$th user at the BS is obtained as
\vspace{-1mm}
\begin{equation}
{\hat \vx}_k = \argmax_{\vs \in \sm} \ p_{k}(\vs).
\label{argprobs}
\end{equation}

\vspace{-3mm}
The non-zero entry in ${\hat \vx}_k$ and its index are then demapped to 
obtain the information bits of the $k$th user. The algorithm listing is 
given in {\bf Algorithm \ref{mpsm}}.

{\em Complexity}:
From (\ref{mu}), (\ref{sigma}), and (\ref{pki}), we see that the total 
complexity of the MPD-SM algorithm is $O(NK|\sm|)$. This complexity is 
less than the MMSE detection complexity of $O(N^2Kn_t)$. Also, the 
computation of double summation in (\ref{mu}) and (\ref{sigma}) can 
further be simplified by using FFT, as the double summation can be 
viewed as a convolution operation.

\begin{algorithm}[t]
{\small
\SetLine
\KwIn{${\bf y}$, ${\bf H}$, $\sigma^2$}
{\bf Initialize}: $p_{ki}^{(0)}(\vs)\gets1/|\sm|$, $\forall i,k,\vs$

\For{$t = 1 \to {\textit Number\_of\_iterations}$ }{
\For{$i = 1 \to N$ }{
\For{$j = 1 \to K$ }{
$\tilde\mu_{ij}\gets\sum\limits_{\vs\in\sm}p_{ji}^{(t-1)}(\vs)s_{l_s}
H_{i,(j-1)n_t+l_s}$
}

$\mu_{i}\gets\sum\limits_{j=1}^{K} \tilde\mu_{ij}$

$\sigma_{i}^2\gets\scriptstyle{\sum\limits_{j=1}^{K} \sum\limits_{\vs\in\sm}
p_{ji}^{(t-1)}(\vs)\big|s_{l_s}H_{i,(j-1)n_t+l_s}\big|^2-\big|\tilde\mu_{ij}\big|^2 
+\sigma^2}$

\For{$k = 1 \to K$ }{
$\mu_{ik}\gets\mu_{i}-\tilde\mu_{ik}$ 

$\sigma_{ik}^2\gets\scriptstyle{\sigma_{i}^2-\sum\limits_{\vs\in\sm}
\hspace{-3mm} p_{ki}^{(t-1)}(\vs)\big|s_{l_s}H_{i,(k-1)n_t+l_s}\big|^2+\big|\tilde\mu_{ik}
\big|^2}$

}
}

\For{$k = 1 \to K$ }{
\ForEach{$\vs \in \sm$}{
$\ln(p_{k}(\vs))\gets\scriptstyle{C_k-\sum\limits_{m=1}^N
\frac{\big|y_m-\mu_{mk}-{\bf h}_{m,[k]}\vs\big|^2}{2\sigma^2_{mk}}}$

$C_k$ is a normalizing constant.
}
\For{$i = 1 \to N$ }{
\ForEach{$\vs \in \sm$}{
$\tilde p_{ki}(\vs)\gets\scriptstyle{ \ln(p_{k}(\vs))+
\ln(\sigma_{ik})+\frac{\big|y_i-\mu_{ik}-
{\bf h}_{i,[k]}\vs\big|^2}{2\sigma^2_{ik}}}$

$p_{ki}^{(t)}(\vs)=(1-\delta)\exp(\tilde p_{ki}^{(t)}(\vs)) + 
\delta p_{ki}^{(t-1)}(\vs)$
}
}
}

}
\KwOut{
$p_{k}(\vs)$ as per (\ref{probs}) and ${\hat \vx}_k$ as per (\ref{argprobs}), 
$\forall k$
}
\label{mpsm}
\vspace{3mm}
\caption{Listing of the proposed MPD-SM algorithm.}
}
\end{algorithm}

\begin{figure}
\hspace{-4mm}
\includegraphics[width=3.75in,height=2.85in]{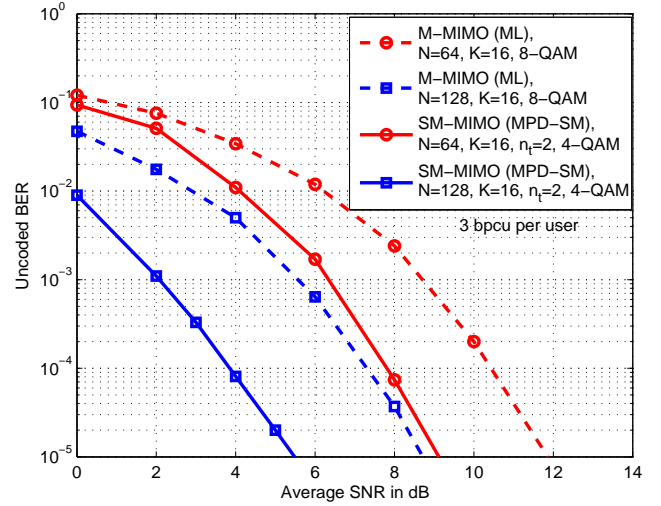}
\vspace{-8mm}
\caption{BER performance of multiuser SM-MIMO ($n_t=2$, $n_{rf}=1$,
4-QAM) using MPD-SM algorithm and massive MIMO ($n_t=1$, $n_{rf}=1$, 8-QAM) 
with sphere decoding, at 3 bpcu per user, $K=16$, $N=64,128$. }
\label{sm3bps}
\vspace{-2mm}
\end{figure}

\begin{figure}
\hspace{-4mm}
\includegraphics[width=3.75in,height=2.85in]{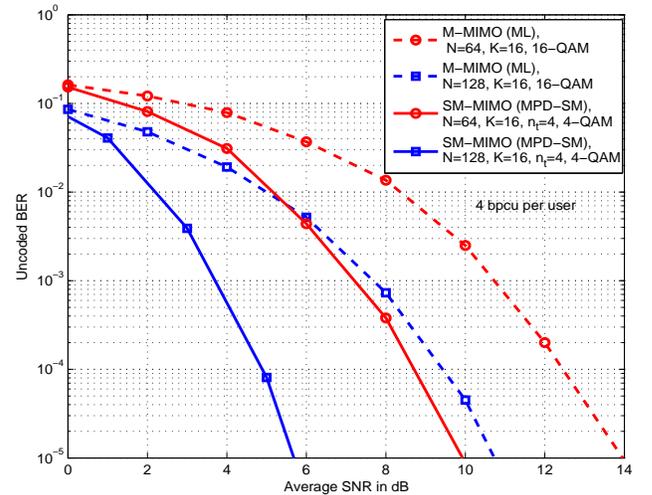}
\vspace{-8mm}
\caption{BER performance of multiuser SM-MIMO ($n_t=4$, $n_{rf}=1$,
4-QAM) using MPD-SM algorithm and massive MIMO ($n_t=1$, $n_{rf}=1$, 16-QAM) 
with sphere decoding, at 4 bpcu per user, $K=16$, $N=64,128$. }
\label{sm4bps}
\end{figure}

{\em Performance}:
We evaluated the performance of multiuser SM-MIMO using the proposed
MPD-SM algorithm and compared it with that of massive MIMO with ML 
detection (using sphere decoder) for the same spectral efficiency with 
$K=16$ and $N=64,128$. It is noted that in both SM-MIMO and massive MIMO 
systems, the number of transmit RF chains at each user is $n_{rf}=1$. For 
SM-MIMO, we consider the number of transmit antennas at each user to be 
$n_t=2,4$. Figure \ref{sm3bps} shows the performance comparison between 
SM-MIMO with ($n_t=2$, 4-QAM) and massive MIMO\footnote{In all the figures, 
massive MIMO is abbreviated as M-MIMO.} 
with ($n_t=1$, 8-QAM), both having 3 bpcu per user.  From Fig. \ref{sm3bps}, 
we can see that SM-MIMO outperforms massive MIMO by several dBs. 
For example, at a BER of $10^{-3}$, SM-MIMO has a 2.5 to 3.5 dB SNR advantage 
over massive MIMO. In Fig. \ref{sm4bps}, we observe a performance advantage 
of about 3 to 4 dB in favor of SM-MIMO with ($n_t=4$, 4-QAM) 
compared to massive MIMO with ($n_t=1$, 16-QAM), both at 4 bpcu per user. 
This SNR advantage in favor of SM-MIMO can be explained as follows. Since SM-MIMO 
conveys information bits through antenna indices in addition to carrying bits on 
QAM symbols, SM-MIMO can use a smaller-sized QAM compared to that
used in massive MIMO to achieve the same spectral efficiency, and a 
small-sized QAM is more power efficient than a larger one. 

\section{Local Search Detection for SM-MIMO}
\label{sec4}
In this section, we propose another algorithm for SM-MIMO detection.
The algorithm is based on local search. The algorithm finds a local 
optimum (in terms of ML cost) as the solution through a local 
neighborhood search. We refer to this algorithm as LSD-SM
(local search detection for spatial modulation) algorithm. A key to
the LSD-SM algorithm is the definition of a neighborhood suited for
SM. This is important since SM carries information bits in the antenna
indices also. 

{\em Neighborhood definition:}
For a given vector $\vx\in \sm^K$, we define the neighborhood ${\cal N}(\vx)$ 
to be the set of all vectors in $\sm^K$ that differ from the vector $\vx$ in 
either one spatial index position or in one modulation symbol. That is,
a vector $\vw$ is said to be a neighbor of $\vx$ if and only if 
$\vw_k \in\{\sm\setminus\vx_k\}$ for exactly one $k$, and $\vw_k=\vx_k$ 
for all other $k$, i.e., the neighborhood ${\cal N}(\vx)$ is given by
\begin{equation}
\hspace{-0mm}
{\cal N}(\vx)\Define\big\{\vw | \vw\in\sm^K, \vw_k\neq\vx_k \text{ for exactly one } k \big\}, \hspace{-4mm}
\label{ngbh}
\end{equation}
where $\vx_k, \vw_k \in \sm$ and $k\in{1, 2, \cdots, K}$. Thus the size 
of this neighborhood is given by $|{\cal N}(\vx)|=(|\sm|-1)K$. 

For example, consider $K=2$, $n_t=2$, and BPSK (i.e., $\sa=\{\pm1\}$).
We then have 

\[
{\small {\mathbb S}_{2,\mbox{\tiny BPSK}}=\Bigg\{
\begin{bmatrix} +1 \\ 0 \end{bmatrix},
\begin{bmatrix} -1 \\ 0 \end{bmatrix},
\begin{bmatrix} 0 \\ +1 \end{bmatrix},
\begin{bmatrix} 0 \\ -1 \end{bmatrix}\Bigg\}},
\]

and 

{\small
${\cal N}\left(\begin{bmatrix} +1 \\ 0 \\ 0 \\-1\end{bmatrix}\right)
\hspace{-1mm}=\hspace{-1mm}\left\{
\begin{bmatrix} -1 \\ 0 \\ 0 \\-1\end{bmatrix},
\begin{bmatrix} 0 \\ +1 \\ 0 \\-1\end{bmatrix},
\begin{bmatrix} 0 \\ -1 \\ 0 \\-1\end{bmatrix},
\begin{bmatrix} -1 \\ 0 \\ 0 \\+1\end{bmatrix},
\begin{bmatrix} -1 \\ 0 \\ -1\\ 0\end{bmatrix},
\begin{bmatrix} -1 \\ 0 \\ +1\\ 0\end{bmatrix}\right\}$.
}

{\em LSD-SM algorithm:}
The LSD-SM algorithm for SM-MIMO detection starts with an initial
solution vector $\hat{\vx}^{(0)}$ as the current solution. For 
example, $\hat{\vx}^{(0)}$ can be the MMSE solution vector 
$\hat{\vx}_{\tiny{\mbox{MMSE}}}$. Using the neighborhood definition 
in (\ref{ngbh}), it considers all the neighbors of $\hat{\vx}^{(0)}$ 
and searches for the best neighbor with least ML cost which also has a 
lesser ML cost than the current solution. If such a neighbor is found, 
then it declares this neighbor as the current solution. This completes
one iteration of the algorithm. This process is repeated for multiple
iterations till a local minimum is reached (i.e., no neighbor better 
than the current solution is found). The vector corresponding to the
local minimum is declared as the final output vector ${\hat \vx}$. The 
non-zero entry in the $k$th user's sub-vector in ${\hat \vx}$ and its 
index are then demapped to obtain the information bits of the $k$th user.

{\em Multiple restarts:} The performance of the basic LSD-SM algorithm 
in the above can be further improved by using multiple restarts, where the 
LSD-SM algorithm is run several times, each time starting with a different 
initial solution and declaring the best solution among the multiple runs. 
The proposed LSD-SM algorithm with multiple restarts is listed in 
{\bf Algorithm \ref{algo2}}. 

\begin{algorithm}      
\caption{Listing of the proposed LSD-SM algorithm with multiple restarts.}
   \begin{algorithmic} [1] 
      \STATE $\mathbf{Input: y, H}$, $r$: no. of restarts
      \FOR{$j$ = 1 \TO $r$ } 
      \STATE compute $\mathbf{c}^{(j)}$ \quad (initial vector at $j$th restart)
 \STATE find $\mathcal{N}(\mathbf{c}^{(j)})$
 \STATE $\mathbf{z}^{(j)}$ = $ \underset{\mathbf{q}\in\mathcal{N}(\mathbf{c}^{(j)})}
  {\mathrm{argmin}} ~\| \mathbf{y-Hq}\|^2$
\IF{ $\| \mathbf{y-H\mathbf{z}}^{(j)}\|^2 < \| \mathbf{y-H\mathbf{c}}^{(j)}\|^2$ }

   \STATE  $\mathbf{c}^{(j)}$ =  $\mathbf{z}^{(j)}$ 
   \STATE  goto step 4
 \ELSE
  \STATE  $\hat{\mathbf{x}}^{(j)}$ =  $\mathbf{c}^{(j)}$ 
 \ENDIF 
 \ENDFOR 
 \STATE $i$ = $\underset{ 1 \leq j \leq r}{\mathrm{argmin}} 
  \ \| \mathbf{y-H}\hat{\mathbf{x}}^{(j)}\|^2$
  
 \STATE $\mathbf{Output}$ : ${\hat \vx} = \hat{\mathbf{x}}^{(i)}$
\vspace{3mm}
\end{algorithmic}
\label{algo2}
\end{algorithm}        

{\em Complexity:}
The LSD-SM algorithm complexity consists of two parts. The first part 
involves the computation of the initial solution. The complexity for
computing the MMSE initial solution is $\mathcal{O}(Kn_tN^2)$. The 
second part involves the search complexity, where, in order to compute
the ML cost, we require to compute $(i)$ $\mh^H\mh$ which has  
$\mathcal{O}(K^2n_t^2N)$ complexity, and $(ii)$ $\mh^H\vy$ which has 
$\mathcal{O}(Kn_tN)$ complexity. In addition, the complexity per
iteration and the number of iterations to reach the local minima 
contribute to the search complexity, where the search complexity per 
iteration is $O(K|\sm|)$. 

{\em Reducing the search complexity}: From the above discussion on the complexity 
of the LSD-SM algorithm, we saw that the computation of the ML cost requires a 
complexity of order $O(K^2n_t^2N)$ which is greater than the MMSE complexity of
$O(Kn_tN^2)$ for systems with $Kn_t>N$, i.e., with loading factor $\alpha>1/n_t$.
We propose to reduce the search complexity by the following
method, which consists of the following three parts:
\begin{enumerate}
\item
The channel gain matrix $\mh$ can be written as 
$\mh=[\vh_1 \ \vh_2 \ \cdots \ \vh_{Kn_t}]$, where $\vh_i$ is the $i$th column
of $\mh$, which is a $N\times1$ column vector. Before we start the search process
in the LSD-SM algorithm, compute the set of vectors
$\mathbb{J}\Define\{\vh_i s\}_{\forall s\in\sa, \forall i\in 1,2,\cdots,Kn_t}$.
The complexity of this computation is $O(|\sa|Kn_tN)$.
\item
Compute the vector $\vz^{(0)}$, which is defined as
\begin{equation}
\vz^{(0)}\Define\vy-\mh\hat{\vx}^{(0)}=\vy-\sum_{k=1}^{K} \hat{x}^{(0)}_{l_k}\vh_{(k-1)n_t+j_k},
\label{init}
\end{equation}
where the terms $\hat{x}^{(0)}_{l_k}\vh_{(k-1)n_t+j_k}$ belong to
$\mathbb{J}$ which is precomputed. 
The computation of $\vz^{(0)}$ requires a complexity of $O(KN)$.
\item
Because of the way the neighborhood is defined, every neighbor of $\vz^{(0)}$ can be 
computed from $\vz^{(0)}$ by exactly adding a single vector from $\mathbb{J}$ and 
subtracting another vector from $\mathbb{J}$. Thus the complexity of computing the 
ML cost of every neighbor is $O(N)$.
\end{enumerate}
In this method, the total number of operations performed for the search is
$|\sa|Kn_tN+K(N+1)+(2N-1)+K(|\sa|n_t-1)(4N-1)T$, where $T$ is the number of 
iterations performed to reach the local minima which depends on the 
transmit vector and the operating SNR ($T$ is determined through simulations).
Therefore, the total complexity of the algorithm in this method is given
by $O(|\sa|Kn_tNT)$, whereas, the total complexity without search 
complexity reduction is $O(K^2n_t^2N)$. 

\begin{figure}
\hspace{-4mm}
\includegraphics[width=3.75in,height=2.85in]{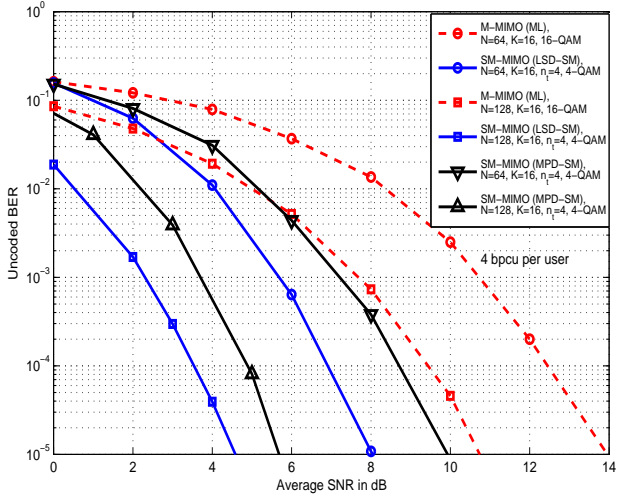}
\vspace{-7mm}
\caption{BER performance of multiuser SM-MIMO ($n_t=4$, $n_{rf}=1$, 
4-QAM) using LSD-SM and MPD-SM algorithms, and massive MIMO ($n_t=1$, $n_{rf}=1$, 
16-QAM) using sphere decoding, at 4 bpcu per user, $K=16$, $N=64,128$. }
\label{sm4bpslas}
\vspace{-4mm}
\end{figure}

{\em Performance}:
We evaluated the performance of multiuser SM-MIMO using the proposed LSD-SM 
algorithm and compared it with that of massive MIMO using ML detection for 
the same spectral efficiency. Figure \ref{sm4bpslas} shows the performance 
comparison between SM-MIMO with ($n_t=4$, 4-QAM) and massive MIMO with 
($n_t=1$, 16-QAM), both having 4 bpcu per user. For SM-MIMO, detection 
performance of both LSD-SM (presented in this section) and MPD-SM (presented 
in the previous section) are shown. In LSD-SM, the number of restarts used
is $r=2$. The initial vectors used in the first and second restarts are MMSE 
solution vector and random vector, respectively. For massive MIMO, ML detection 
performance using sphere decoder is plotted. It can be seen that SM-MIMO using 
LSD-SM and MPD-SM algorithms outperform massive MIMO using sphere decoding. 
Specifically, SM-MIMO using LSD-SM performs better than massive MIMO by about 
5 dB at $10^{-3}$ BER. Also, comparing the performance of LSD-SM and MPD-SM 
algorithms in SM-MIMO, we see that LSD-SM performs better than MPD-SM by about 
1 dB at $10^{-3}$ BER. 

{\em Hybrid MPD-LSD-SM detection}:
The LSD-SM algorithm proposed in this section offers good performance but 
has higher complexity due to the requirement of the initial MMSE solution 
vector. The high complexity of MMSE is due to the need for matrix inversion. 
We can overcome this need for MMSE computation by using a hybrid detection
scheme. In the hybrid detection scheme, we first run the MPD-SM algorithm 
(proposed in the previous section) and the output of the MPD-SM algorithm
is fed as the initial solution vector to the LSD-SM algorithm (proposed in
this section). We refer to this hybrid scheme as the `MPD-LSD-SM' scheme.
The MPD-LSD-SM scheme does not need the MMSE solution and hence avoids 
the associated matrix inversion. 

\begin{figure}
\hspace{-5mm}
\includegraphics[width=3.75in,height=2.85in]{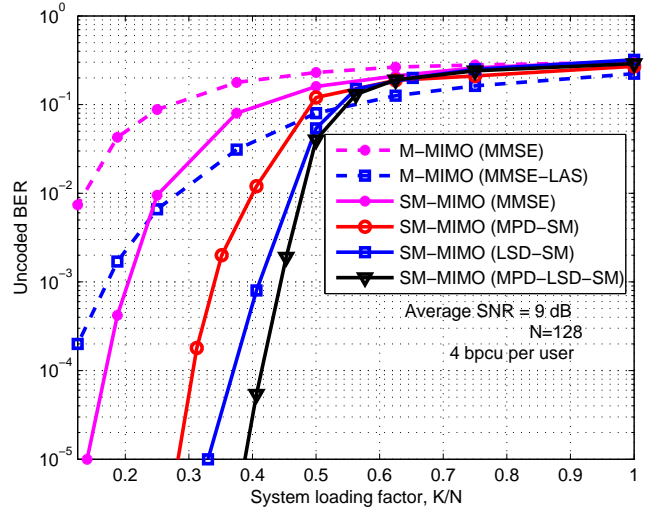} 
\vspace{-6mm}
\caption{BER performance of SM-MIMO ($n_t=4$, $n_{rf}=1$, 4-QAM) and 
massive MIMO ($n_t=1$, $n_{rf}=1$, 16-QAM) as a function of system loading 
factor, $\alpha$. $N=128$, SNR = 9 dB, and 4 bpcu per user. }
\label{alpha_p} 
\vspace{-4mm}
\end{figure}

\begin{figure}
\hspace{-5mm}
\includegraphics[width=3.75in,height=2.85in]{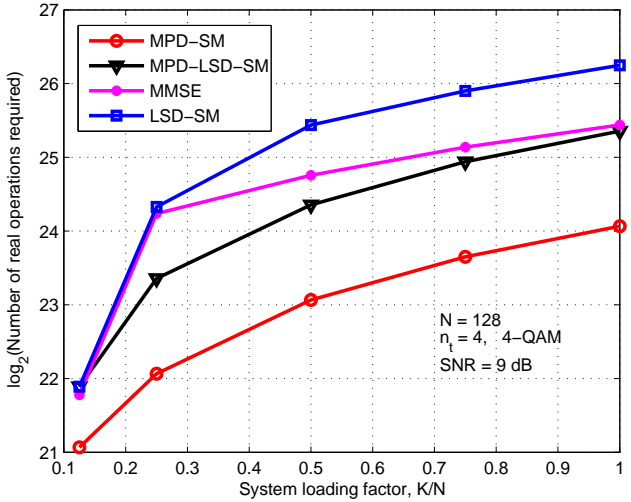} 
\vspace{-7mm}
\caption{Complexity comparison between MMSE, MPD-SM, LSD-SM and hybrid MPD-LSD-SM 
detection algorithms in multiuser SM-MIMO as a function of system loading factor, 
$\alpha$. $N=128$, $n_t=4$, $n_{rf}=1$, 4-QAM, 4 bpcu per user and SNR = 9 dB.}
\label{alpha_c} 
\vspace{-5mm}
\end{figure}

{\em Performance as a function of loading factor:}
In Fig. \ref{alpha_p}, we compare the performance of SM-MIMO (with $n_t=4$,
$n_{rf}=1$, 4-QAM) and massive MIMO (with $n_t=n_{rf}=1$, 16-QAM), both at
4 bpcu per user, as a function of system loading factor $\alpha$, at an
average SNR of 9 dB. For SM-MIMO, the detectors considered are MMSE, MPD-SM,
LSD-SM, and the hybrid MPD-LSD-SM. The detectors considered for massive MIMO are 
MMSE detector and MMSE-LAS detector in \cite{lasx},\cite{lasy} with 2 restarts. 
From Fig \ref{alpha_p}, we observe that SM-MIMO performs significantly better 
than massive MIMO at low to moderate loading factors. For the same SM-MIMO 
system settings, we show the complexity plots for various SM-MIMO detectors at 
different loading factors in 
Fig. \ref{alpha_c}. It can be seen that the proposed MPD-SM detector has less 
complexity than MMSE detector; yet, MPD-SM detector outperforms MMSE detector 
(as can be seen in Fig. \ref{alpha_p}). The proposed LSD-SM detector performs 
better than the MPD-SM detector with some additional computational complexity 
(as can be seen in Fig. \ref{alpha_c}). Among the considered detection schemes,
the hybrid MPD-LSD-SM detection scheme gives the best performance with near-MMSE 
complexity.

{\em Performance for same spectral efficiency and QAM size:}
We note that if both spectral efficiency and QAM size are to be kept
same in SM-MIMO and massive MIMO, then the number of spatial streams
per user in massive MIMO has to increase. For example, SM-MIMO can 
achieve 4 bpcu per user with 4-QAM using $n_t=4$ and $n_{rf}=1$. 
Massive MIMO can achieve the same spectral efficiency of 4 bpcu per
user using one spatial stream (i.e., $n_t=n_{rf}=1$) with 16-QAM. 
But to achieve the same spectral efficiency using 4-QAM in massive
MIMO, we have to use $n_t=n_{rf}=2$, i.e., two spatial streams per
user with 4-QAM on each stream are needed. This increase in number 
of spatial streams per user increases the spatial interference.

\begin{figure}
\hspace{-5mm}
\includegraphics[width=3.75in,height=2.85in]{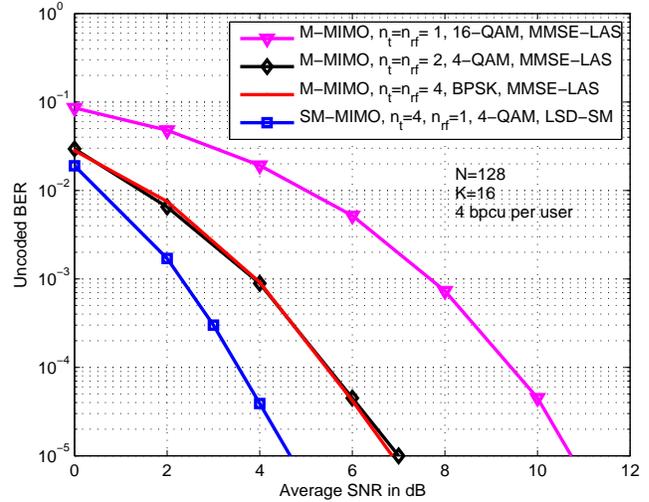} 
\vspace{-7mm}
\caption{BER performance of SM-MIMO with ($n_t=4$, $n_{rf}=1$, 4-QAM), 
massive MIMO with ($n_t=n_{rf}=1$, 16-QAM), massive MIMO with
($n_t=n_{rf}=2$, 4-QAM), and massive MIMO with 
($n_t=n_{rf}=4$, BPSK) for $K=16$, $N=128$, 4 bpcu per user. }
\label{new_fig} 
\vspace{-5mm}
\end{figure}

The effect of increase in number of spatial streams per user
in massive MIMO for the same spectral efficiency on the performance 
is illustrated in Fig. \ref{new_fig} for $K=16$ and $N=128$. 
In Fig. \ref{new_fig}, we compare the performance of 
the following four systems with the same spectral efficiency of 4 bpcu
per user:
1) SM-MIMO with ($n_t=4$, $n_{rf}=1$, 4-QAM), 2) massive MIMO with 
($n_t=n_{rf}=1$, 16-QAM), 3) massive MIMO with ($n_t=n_{rf}=2$, 4-QAM), 
and 4) massive MIMO with ($n_t=n_{rf}=4$, BPSK). 
It can be seen that among the four systems considered in Fig. \ref{new_fig},
SM-MIMO performs the best. This is because massive MIMO loses performance
because of higher-order QAM or increased spatial interference from
increased number of spatial streams per user. 

\section{Conclusions}
\label{sec5}
We proposed low complexity detection algorithms for 
large-scale SM-MIMO systems. These algorithms, based on message passing
and local search, scaled well in complexity and achieved very good 
performance. An interesting observation from the simulation results is
that SM-MIMO outperforms massive MIMO by several dBs for the same spectral 
efficiency. The SNR advantage of SM-MIMO over massive MIMO is attributed 
to the following reasons: $(i)$
because of the spatial index bits, SM-MIMO can use a lower-order QAM
alphabet compared to that in massive MIMO to achieve the same spectral
efficiency, and $(ii)$ for the same spectral efficiency and QAM size,
massive MIMO will need more spatial streams per user which leads to
increased spatial interference.
With such performance advantage at low RF hardware complexity,
large-scale multiuser SM-MIMO is an attractive technology for next
generation wireless systems and standards like 5G and HEW (high
efficiency WiFi).

\vspace{-2mm}
\bibliographystyle{ieeetr} 

\end{document}